\documentclass[12pt]{iopart}

\usepackage{graphicx}
\usepackage[english]{babel}
\usepackage{subfigure}
\usepackage{wrapfig}

\begin{document}
\newcommand{\mus}{$\mu$s }

\title[Solitary magnetic perturbations at the ELM onset]{Solitary magnetic perturbations at the ELM onset}
\author{RP Wenninger$^1$, H Zohm$^2$, JE Boom$^3$, A Burckhart$^2$, MG Dunne$^4$, R Dux$^2$, T Eich$^2$,  R Fischer$^2$,  C Fuchs$^2$, M Garcia-Munoz$^2$, V Igochine$^2$, M H\"olzl$^2$, NC Luhmann Jr$^5$, T Lunt$^2$, M Maraschek$^2$, HW M\"uller$^2$, HK Park$^6$, PA Schneider$^2$, F Sommer$^2$, W Suttrop$^2$, E Viezzer$^2$ and the ASDEX Upgrade Team}
\address{$_1$Universit\"atssternwarte der Ludwig-Maximilians-Universit\"at, M\"unchen, Germany}
\address{$_2$Max-Planck-Institut f\"ur Plasmaphysik, EURATOM Association, Garching, Germany}
\address{$_3$FOM Institute DIFFER - Dutch Institute for Fundamental Energy Research, Association EURATOM/FOM, The Netherlands}
\address{$_4$Department of Physics, University College Cork, Association Euratom-DCU, Cork, Ireland}
\address{$_5$University of California, Davis, CA 95616, USA}
\address{$_6$POSTECH, Pohang, Gyeongbuk, 790-784, Korea}

\ead{ronald.wenninger@ipp.mpg.de}

\begin{abstract}
Tokamak H-mode plasmas frequently exhibit edge localised modes (ELMs). ELMs allow maintaining sufficient plasma purity and thus enable stationary H-mode. On the other hand in a future device ELMs may cause divertor power flux densities far in excess of tolerable material limits. The size of the energy loss per ELM is determined by saturation effects in the non-linear phase of the ELM, which at present is hardly understood. ASDEX Upgrade is now equipped with a set of fast sampling diagnostics, which is well suited to investigate the chain of events around the ELM crash with appropriate temporal resolution ($\leq10\mu s$).\\
Solitary magnetic perturbations (SMPs) are identified as dominant features in the radial magnetic fluctuations below 100kHz. They are typically observed close ($\pm100\mu s$) to the onset of pedestal erosion. SMPs are field aligned structures rotating in the electron diamagnetic drift direction with perpendicular velocities of about $10km/s$. A comparison of perpendicular velocities suggests that the perturbation evoking SMPs is located at or inside the separatrix. Analysis of very pronounced examples showed that the number of peaks per toroidal turn is 1 or 2, which is clearly lower than corresponding numbers in linear stability calculations. In combination with strong peaking of the magnetic signals this results in a solitary appearance resembling modes like palm tree modes, edge snakes or outer modes. This behavior has been quantified as \textit{solitariness} and correlated to main plasma parameters.\\
SMPs may be considered as a signature of the non-linear ELM-phase originating at the separatrix or further inside. Thus they provide a handle to investigate the transition from linear to non-linear ELM phase. By comparison with data from gas puff imaging processes in the non-linear phase at or inside the separatrix and in the scrape-off-layer (SOL) can be correlated. A connection between the passing of an SMP and the onset of radial filament propagation has been found. Eventually the findings related to SMPs may contribute to a future quantitative understanding of the non-linear ELM evolution.

\end{abstract}
\section{Introduction}
In the high confinement regime (H-mode) \cite{WAGNER82A} tokamak plasmas exhibit edge localized modes (ELMs) \cite{ZOHM96A}. ELMs are plasma edge instabilities of bursty nature, which enhance particle transport and thus allow maintaining sufficient plasma purity and enable stationary H-mode. On the other hand in a future device ELMs may cause energy deposition far in excess of tolerable material limits \cite{FEDERICI03B}. Hence the size of energy loss per ELM has to be controlled.\\
This size is determined by saturation effects in the non-linear phase of the ELM, which at present is hardly understood. The central idea of this work is to reconstruct aspects of the non-linear evolution of spontaneously occurring ELMs on the basis of data from several fast sampling diagnostics regarding the exact position of these. The main focus of the analysis is on magnetic perturbations, which have served in numerous previous studies as an indicator of the intensity of MHD related activity. We identify solitary magnetic perturbations (SMPs) (i.e. featuring isolated sets of peaks and dips) frequently to be the dominant features in the radial magnetic fluctuations below 100kHz.\\
Our findings show that SMPs are features of the non-linear ELM phase, as they are different to structures found by linear stability calculations \cite{SNYDER09A}. We observe them appearing usually close to ($\pm100\mu s$) the onset of the pedestal crash.\\
In section \ref{sec:exp_arr} we describe the set of analyzed discharges, main employed diagnostics and central analysis tools. Section \ref{sec:ELM_dyn} provides information on the typical ELM related dynamics of the pedestal, plasma stored energy, divertor $D_\alpha$ radiation and divertor current. The latter is used to generate an ELM onset marker. After this preparatory part, in section \ref{sec:SMPs} SMPs, the main focus of this publication, are addressed. After introducing SMPs, a quantification of the level of solitary appearance is presented, which is then correlated to other plasma parameters for an extensive set of ELMs. The major part of section \ref{sec:SMPs} presents information on the structure of SMPs. In section \ref{sec:gpi} results from gas puff imaging of ELM filaments are reported and correlated to signatures of SMPs. Finally in section \ref{sec:sum_dis} main findings are summarized and discussed.\\
\section{Experimental arrangement}
\label{sec:exp_arr}
\subsection{Analyzed discharges}
\label{subsec:ana_disch}
Table 1 summarizes a number of plasma parameters of the discharges that are analyzed in detail in this paper. The abbreviations in table 1 correspond to (\#) number of the discharge, ($\Delta t$) time interval of flat top phase used for data analysis, (Conf) configuration (LSN / USN: lower single null / upper single null), ($I_P$) plasma current, ($B_T$) toroidal magnetic field, ($\delta_{up}$ and $\delta_{low}$) upper and lower triangularity at the separatrix, ($q_{95}$) safety factor at the surface of 95\% of the poloidal magnetic flux at the separatrix, ($W_{mhd}$) plasma stored energy obtained from equilibrium reconstruction, ($n_{edge}$) line averaged density from interferometry edge channel, ($P_{NBI}$) neutral beam input power, ($P_{ICRH}$) ion cyclotron resonance heating input power and ($P_{ECRH}$) electron cyclotron resonance heating input power. All quantities are averaged over the relevant time interval.\\
\begin{table}[ht] \footnotesize
\label{tab}
\caption{Overview of the main parameters of the discharges analyzed in this paper}
\centering
\begin{tabular}{c c c c c c c c c c c c c c c}
\hline\hline
\#	& $\Delta t$ & Conf & $I_P$ & $B_T$ & $\delta_{up}$ & $\delta_{low}$ & $q_{95}$ & $W_{mhd}$ & $n_{edge}$ & $P_{NBI}$ & $P_{ICRH}$ & $P_{ECRH}$\\
 &	[s]& & [MA] & [T] & [1] & [1] & [1] & [MJ] & [$10^{19} m^{-3}$] & [MW] & [MW] & [MW]\\
\hline
24059 & 3.7 - 3.8 & LSN & 1.0 & -2.5 & 0.09 & 0.38 & -4.7 & 0.48 & 7.1 & 4.8 & 1.2 & 0\\
25764 & 1.6 - 1.9 & LSN & 1.0 & -2.4 & 0.10 & 0.42 & -4.4 & 0.86 & 5.3 & 7.5 & 0 & 0.9\\
26299 & 1.8 - 2.8 & LSN & 1.0 & -2.5 & 0.05 & 0.44 & -4.0 & 0.63 & 6.8 & 7.4 & 0 & 0.9\\
26324 & 2.5 - 3.5 & LSN & 1.0 & -2.5 & 0.26 & 0.48 & -4.4 & 0.77 & 7.3 & 7.7 & 0 & 0.8\\
26510 & 1.4 - 1.5 & LSN & 0.8 & -2.0 & 0.09 & 0.39 & -4.3 & 0.39 & 6.7 & 4.8 & 0 & 0\\
26703 & 2.3 - 2.4 & USN & 1.0 & -2.5 & 0.34 & 0.15 & -4.2 & 0.62 & 10.5 & 7.4 & 0 & 0.8\\
26704 & 1.7 - 1.8 & USN & 1.0 & 2.5 & 0.38 & 0.18 & 4.4 & 0.86 & 7.2 & 7.3 & 0 & 0.8\\
26764 & 4.0 - 5.0 & LSN & 0.8 & -2.4 & 0.04 & 0.45 & -5.3 & 0.52 & 4.2 & 5.4 & 0 & 0.9\\
27082 & 2.3 - 2.4 & LSN & 1.0 & -2.5 & 0.09 & 0.36 & -4.6 & 0.55 & 7.0 & 4.9 & 0 & 2.1\\
27112 & 1.9 - 2.0 & LSN & 1.0 & -2.4 & 0.07 & 0.35 & -4.3 & 0.47 & 7.3 & 4.4 & 0 & 0\\
\hline
\end{tabular}
\end{table}
\subsection{Fast sampling diagnostics}
\begin{figure}[ht]
  \centering
  \includegraphics[scale=0.5]{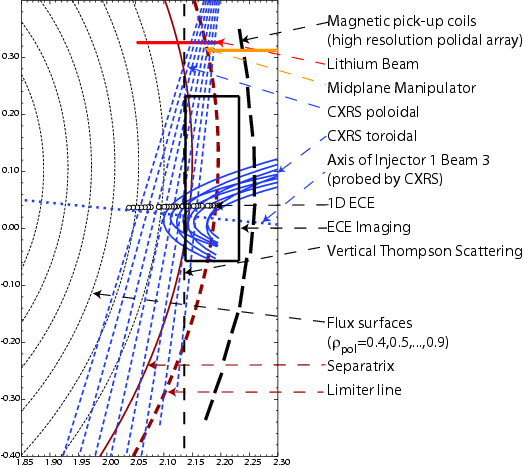}
  \caption{\label{fig:diaggeom2} Outboard part of poloidal cross section of ASDEX Upgrade together with equilibrium and a selection of employed diagnostics.}
\end{figure}

Figure \ref{fig:diaggeom2} provides an overview of the geometry of some of the diagnostics used in this work. Due to the time scales related to ELMs fast sampling diagnostics are of clear advantage. Magnetic perturbations associated with ELMs are a main focus and a starting point of this work. Therefore magnetic probes measuring at a rate of 2MHz are extensively employed. Probes measuring $dB_{rad}/dt$ have mainly been used, as in ASDEX Upgrade these are located close to the plasma (see figure \ref{fig:diaggeom2}) and therefore resolve higher multipole moments of a perturbation structure well. The positions of the probes of two poloidal and one toroidal array can be seen in figure \ref{fig:mapping_trajectories}. Other fast sampling diagnostic systems which we have employed are the electron cyclotron emission imaging (ECEI) system \cite{BOOM11A}, fast framing cameras and the measurement of the divertor current. These will be briefly described in the relevant sections.
\subsection{Analysis tools}
\subsubsection{Frequency band selection}
\begin{figure}[ht]
  \centering
  \includegraphics[scale=0.4]{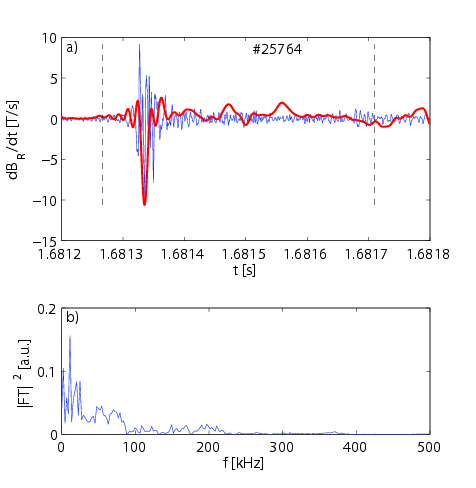}
  \caption{\label{fig:mag_hi_low_decomp3} a) Decomposition of signal of the time derived radial magnetic field during an ELM in discharge 25764 in a high (blue) and low (red) frequency component. b) Squared modulus of the Fourier transformed of the original signal for the time interval confined by the vertical dashed lines in a).}
\end{figure}

The typical ELM signature on pickup coils measuring the radial magnetic field at ASDEX Upgrade is broadband extending from a few kHz to several hundreds of kHz. Figure \ref{fig:mag_hi_low_decomp3}a) shows the decomposition of such a signal into high and low frequency components. To produce the low frequency component a Chebyshev Filter ($f_{pass}=60\textrm{kHz}, f_{stop}=100\textrm{kHz}$) is applied. The resulting low frequency component is subtracted from the original signal to obtain the high frequency component.\\
The high frequency component usually shows a very fast rise close to a prompt onset. This is frequently followed by a decay with an exponential-like envelope. Particularly in the decay phase, high and low frequency components can exhibit quite different dynamics. In the example illustrated in figure \ref{fig:mag_hi_low_decomp3}a) the low frequency component still has relatively high excursions, while the high frequency part has decayed to a level significantly below the peak level. Figure \ref{fig:mag_hi_low_decomp3}b) shows the squared modulus of the Fourier transform of the original signal for the time interval indicated in figure \ref{fig:mag_hi_low_decomp3}a), which is proportional to the power spectral density. This distribution shows highest values for frequencies lower than 50kHz and a clear decay from 80 to 100kHz. From these observations we assume that high and low frequency components are the footprint of at least two different physical processes, which may well be strongly linked.\\
The focus of this analysis is on ELM associated plasma edge phenomena. Analyzing temperature and density measurements obtained by Thomson scattering typically 10 to 20 blobs per full toroidal rotation are found \cite{KURZAN05A}. In another study structures observed by infrared thermography in the divertor have been mapped back to the mid plane and extrapolated to 8 to 20 peaks per toroidal rotation \cite{EICH03B}. Both ranges agree with the corresponding values for filaments observed by gas puff imaging in this work (see section \ref{sec:gpi}). Linear MHD stability analysis finds similar upper limits for the mode number of the most unstable components in peeling-ballooning modes \cite{SNYDER09A}. Recent nonlinear gyrofluid simulations \cite{KENDL10A} suggest that ion temperature gradient driven micro instabilities may also be involved in the mechanism of the ELM. Although this would involve smaller scales ($n \sim 30 - 100$) than described by linear MHD we do not focus on these here. We use 6kHz as an upper limit for the plasma toroidal rotation frequency, as derived from observations (see section \ref{subsec:rad_loc_smp}). This limits the frequency with which the features of interest are expected to pass the probes to below 120kHz. On the basis of these considerations, and in order to simplify tracking of these features over several adjacent diagnostic channels, we use the low pass filter described above for the analysis of data recorded with high time resolution.
\subsubsection{Diagnostic mapping}
\label{subsubsec:diagmap}
\begin{figure}[ht]
  \centering
  \includegraphics[scale=0.4]{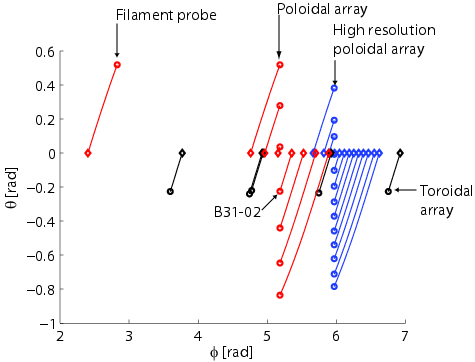}
  \caption{\label{fig:mapping_trajectories} Trajectories of diagnostic mapping procedure for probes measuring time derived radial magnetic field in discharge 25764 at 1.757s from diagnostic reference point (circles) to mapping target point (diamonds). The flux surface with $\rho_{pol}=0.95$ has been used.}
\end{figure}

As described above various diagnostics are available at ASDEX Upgrade, which are capable of temporally resolving ELM associated processes. However these measurements are probing the plasma at an extensive variety of toroidal, poloidal and radial positions. Additionally, some measurements are volume integrated, some are line integrated and some are well localised.\\
An accurate approach to correlate different signals recorded during ELMs must take into account these different measurement positions. Therefore we have developed a method of mapping measurement positions to a common reference surface. We do this on the assumption, that for the investigated structures, the parallel wave number is so small that plasma parameters, like density or temperature, do not change along a field line on the magnetic low field side. This assumption will be justified qualitatively in section \ref{subsec:par_struc}.\\
For line or volume averaged measurements, we identify each diagnostic with a certain diagnostic reference point, which can be regarded as the center of the intersection of the probed area with the area of existence of the investigated feature. For instance, for the  magnetic probe measurement of a certain feature located in the steep gradient region, this diagnostic reference point is taken as the intersection of the shortest connection of the probe and the magnetic axis with an appropriate plasma edge flux surface. From these reference points we trace field lines to a common reference surface. This surface can be a poloidal cross-section ($\phi=const.$) or a surface of constant poloidal angle. In this work field lines are traced from the reference point to the outer mid plane ($\theta=0^\circ$). This particular approach can be applied to measurements inside and outside the separatrix. Figure \ref{fig:mapping_trajectories} shows the mapping trajectories of two poloidal and one toroidal arrays of probes measuring the time derived radial magnetic field component. In the following the toroidal angle of the end points (diamonds) will be called toroidal mapping target angle $\phi_{map}$.
\section{ELM dynamics of typical edge quantities}
\label{sec:ELM_dyn}
\begin{figure}[ht]
  \centering
  \includegraphics[scale=0.4]{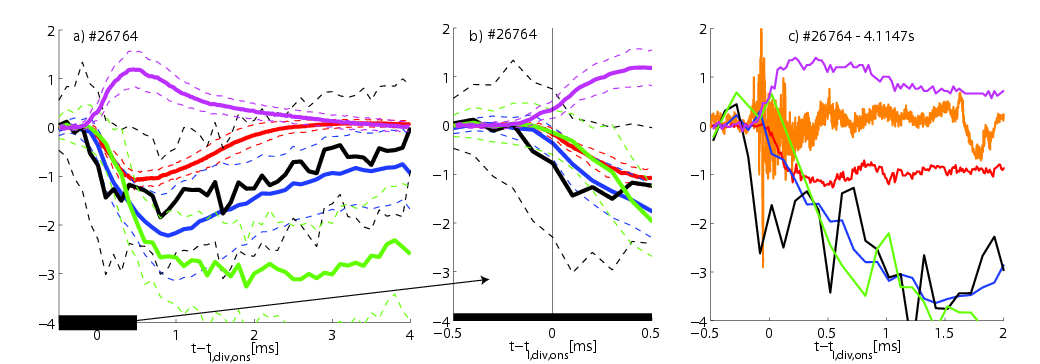}
  \caption{\label{fig:26764_ELM_dyn3} a) Average evolution of scaled (factors in square brackets) and offset plasma quantities during ELMs in discharge 26764 from 4.0 to 5.0s (bold lines): Current to the outer divertor [$10^4$ A] (red), electron temperature at $\rho_{pol}=0.95$ [100eV] (blue), electron density at $\rho_{pol}=0.95$ [$5\times10^{18} m^{-3}$] (black), plasma stored energy [$10^4$ J] (green) and $D_\alpha$ radiation in the outer divertor [a.u.] (magenta). Dashed lines indicate intervals of one standard deviation. b) Close up of a) for an interval of 1ms around $t_{I,div,ons}$. c) Evolution of the same quantities plus time derived magnetic field [10T/s] (orange) measured at a position in the vicinity of ECE and Lithium Beam measurements for a single ELM.}
\end{figure}

ELMs are known to be associated with drops of typical plasma parameters like electron temperature \cite{ZOHM96A} and density \cite{NUNES06A} on the pedestal top and plasma stored energy. Under certain conditions related to the detachment state \cite{PETRIE92A} the $D_\alpha$ radiation in the outer divertor shows a strong increase at the same time. Correlated to these processes, the current $I_{pol,sol,out}$ has a strong rise in absolute value \cite{BURCKHART10A}. This current is the sum of the currents through four shunts connected to the outer divertor tiles with identical toroidal but different poloidal positions in the SOL. As this current can be acquired easily with high temporal resolution, it constitutes a good diagnostic to generate ELM time markers on a regular basis. For each discharge listed in table 1 we have identified the start times $t_{I,div,ons}$ of the rise of $|I_{pol,sol,out}|$.\footnote{An algorithm has been used that detects, after subtraction of a pre-ELM average, a time in the rise phase corresponding to $5-25\%$ of the peak value.}\\
To characterize the basic evolution of a typical ELM we calculate the coherent ELM averages of five quantities: quoted above. Electron temperature and density profiles are obtained via Integrated Data Analysis (IDA) \cite{FISCHE10A} from Electron Cyclotron Emission Spectroscopy and Lithium Beam Diagnostic in combination with Interferometry. We use the IDA values at $\rho_{pol}=0.95$ (pedestal top). The plasma stored energy is obtained from equilibrium reconstruction with 0.1ms time resolution\footnote{For non-circular plasmas integral quantities like $W_{mhd}$ are well recoverable by equilibrium reconstruction \cite{BRAAMS90A}. However, during the most transient phase (about $50$ to $100\mu s$) the exact values of $W_{mhd}$ should be treated with care. In this analysis we focus on the timing and the dynamics of the onset of $W_{mhd} reduction$. This information is considered to be robust.}.\\
In discharge 26764 from 4.0 to 5.0s 71 ELMs are found. For each of the quantities time traces during all ELMs are aligned with respect to $t_{I,div,ons}$ and offset by an average over an initial phase of the investigated time interval. For each time step averages and standard deviations are calculated. Figures \ref{fig:26764_ELM_dyn3}a) and b) show these averages (thick lines) and the intervals of one standard deviation around them (thin lines).\\
First it is important to note that the scatter of the $I_{pol,sol,out}$ trajectories is relatively minor, even though the associated measurements are carried out at one toroidal position. Furthermore one can see that the rise of $|I_{pol,sol,out}|$ begins\\
- at about the same time as the drop of electron temperature at the pedestal top\\
- about 0.2ms later than as the drop of electron density at the pedestal top\footnote{Note the relatively large standard deviation for the density!}\\
- in sync with the drop in plasma stored energy and\\
- slightly later than the onset of $D_\alpha$ radiation.\\
In summary $t_{I,div,ons}$ yielded from $I_{pol,sol,out}$ is a suitable marker ($\pm100\mu s$) for the onset of pedestal erosion.
\section{Observation of solitary magnetic perturbations at the ELM onset}
\label{sec:SMPs}
\begin{figure}[ht]
  \centering
  \includegraphics[scale=0.5]{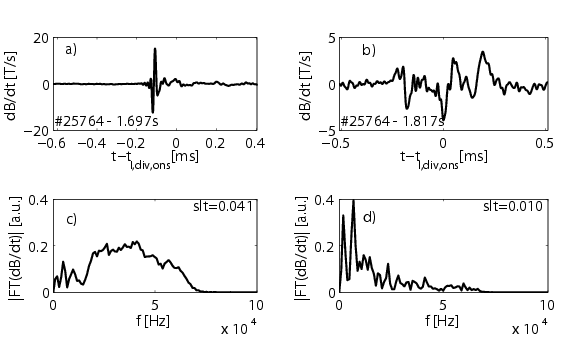}
  \caption{\label{fig:25764_2_elms_sol_comp2} Comparison of magnetic signatures of ELMs in discharge 25764: a) and b) Time trace of time derivative of radial magnetic field of two ELMs at 1.697s and 1.817s. c) and d) Fourier spectra of the waveforms illustrated in a) and b).}
\end{figure}

The low pass filtered magnetic signature has a high level of variation from ELM to ELM. To demonstrate this we compare two ELMs from discharge 25764, which displays very high electron temperatures and low to moderate edge densities. Figures \ref{fig:25764_2_elms_sol_comp2}a) and b) show a comparison of the signatures of the time derived radial magnetic field of two ELMs in this discharge. The strongest perturbation in a) consists of one negative followed by one positive excursion with a total duration in the order of $100\mu s$. In b) the entire perturbation, as well as single peaks, are of significantly longer duration. Furthermore, the number of strong excursions is greater in b) compared to a).\\
The magnetic perturbation in a) has a solitary nature. In comparison, the signal in b) is less solitary with maximal absolute values about 4 times lower. However, even in this case peaked structures can be observed. Therefore, we refer to all magnetic perturbations with such an appearance as \textit{solitary magnetic perturbations (SMPs)}.\\
Besides ELMs, various other H-mode edge instabilities have been reported to show a characteristic signature recorded by magnetic probes. For the edge snake \cite{SOMMER11A}, the Palm Tree Mode \cite{KOSLOWSKI05A} and the Outer Mode \cite{SOLANO2010A} this signature diverges from a sine towards phases of constant signal, periodically interrupted by solitary excursions.\\
\subsection{Quantification and scaling of solitariness}
\label{subsec:quant_sol}
The terminology 'SMP' as introduced above includes an extensive phenomenological range. In order to further discriminate, and to obtain some information on the dependence of the level of solitary appearance on relevant parameters, we develop a quantification of solitariness. The basic idea is that a larger number of Fourier components are necessary for a comparable approximation of a very spiky perturbation, like in figure \ref{fig:25764_2_elms_sol_comp2}a), compared to broad perturbations, as in figure \ref{fig:25764_2_elms_sol_comp2}b). Including also a normalization we define the solitariness \textit{slt} as the ratio of the mean to the maximum of the absolute value of the Fourier Transform of the input signal (here: time derived radial magnetic field) in a certain time interval. Application of this definition to Gaussian input signals yields that \textit{slt} is indirectly proportional to the full width at half maximum (FWHM) of the Gaussian, as long as the FWHM is clearly smaller than the investigated period of time. A delta function has $slt=1$ and a sine has $slt=0$. In the following analysis for the time interval, a period of about 1ms, centered at the maximum absolute value of the input signal in the vicinity of $t_{I,div,ons}$, is chosen.\\
The described definition gives for the examples depicted in figure \ref{fig:25764_2_elms_sol_comp2} solitariness values of 0.041 (a) respectively 0.010 (b). Visual examination of a larger number of ELMs reveals that high \textit{slt} values (e.g. $\ge0.04$) are usually correlated only to a clearly solitary appearance, while low \textit{slt} values can be correlated to any level of solitary appearance. For instance, a very solitary magnetic perturbation can be superimposed by the signature of other, more harmonic MHD-activity leading to a total signal of low solitariness and marginally changed appearance. A clear solitary appearance of a magnetic perturbation is a necessary but not sufficient condition for a high value of the developed quantity \textit{slt}.\\
To gain information on solitariness for a higher number of ELMs, ASDEX Upgrade discharges 26200 to 27200 are systematically analyzed. For each detected ELM \textit{slt} are calculated on the basis of signals from 6 magnetic probes located in the outer mid plane and toroidally spread over $180^\circ$. For each ELM a set of parameters ($I_p$, $W_{mhd}$ \footnote[1]{obtained from equilibrium reconstruction}, average triangularity $\delta_{av}$ \footnotemark[\value{footnote}], $q_{95}$ \footnotemark[\value{footnote}] and density $n_{edge}$ measured by interferometry channel through plasma edge) is also acquired. All parameters are averaged over time intervals starting in the middle of two subsequent ELM onset times and ending at the same point during the next inter ELM period. Table 2 lists the range of these parameters correlated to $slt$ in the investigated data set. Shots suspected to have fringe jumps in the interferometry signal are excluded. ELMs for which the level of the inter ELM magnetic fluctuations exceeds $30\%$ of the peak value are suspected to corrupt due to MHD-activity not related to ELMs and omitted.\\
\begin{table}[tb] \footnotesize
\label{tab}
\caption{Range of parameters correlated to $slt$: The corresponding set of ELMs is described in the text.}
\centering
\begin{tabular}{c c c c}
\hline\hline
Parameter	& Unit & Mean & Standard Deviation\\
\hline
$I_P$          & A   & 0.91   & 0.12\\
$W_{mhd}$      & MJ   & 0.45   & 0.14\\
$\delta_{av}$  & 1   & 0.26   & 0.047\\
$q_{95}$       & 1   & -4.7   & 0.82\\
$n_{edge}$     & $10^{19}$ & 4.2 & 2.1\\
\hline
\end{tabular}
\end{table}
\begin{figure}[ht]
  \centering
  \includegraphics[scale=0.4]{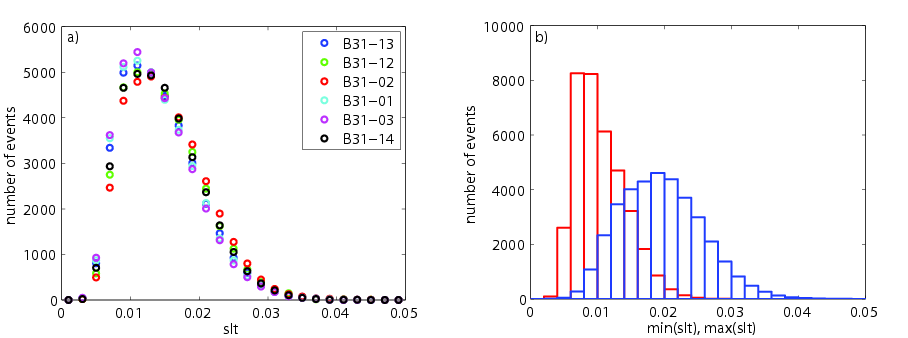}
  \caption{\label{fig:sol_hist2} a) Distribution of \textit{slt} for the selected ELMs (see text) and 6 magnetic probes located in the outer mid plane and toroidally spread over $180^\circ$. b) Histograms of the minimum (red) and maximum (blue) values of \textit{slt} per ELM}
\end{figure}
On the basis of these criteria more than 40000 ELMs are selected. Figure \ref{fig:sol_hist2}a) illustrates the distributions of \textit{slt} for these ELMs for 6 magnetic probes. The distributions are very similar to each other, but are shifted slightly towards higher values for some probes (e.g. B31-02). The peaks of the distributions ($slt\sim0.011$) correspond to a value for \textit{slt} close to the example shown in figure \ref{fig:25764_2_elms_sol_comp2}b). For each ELM the minimum and maximum of \textit{slt} for the 6 probes has been evaluated. The distribution of these extrema are displayed in figure \ref{fig:sol_hist2}b). It is obvious that for single ELMs the values of \textit{slt} obtained via the set of probes has a significant range. This indicates that SMPs may be associated with a significant level of toroidal asymmetry.\\
We address the question under which conditions high solitariness events can be observed in the plasma. As the observation location does not matter in this context we correlate main plasma parameters to the maximum $slt_{max}$ of all solitariness values for each ELM. Attempts to describe $slt_{max}$ by a power law using the parameters $I_p$, $W_{mhd}$, $\delta_{av}$, $q_{95}$ and $n_{edge}$ have not been successful. One possible reason for this is that a clear solitary appearance is not a sufficient condition for a high value of $slt_{max}$. Looking at the correlation of single parameters to $slt_{max}$ yields that the distribution of $slt_{max}$ is pushed towards higher values if $n_{edge}$ is decreased or $|q_{95}|$ or $\delta_{av}$ is increased.\\
Finally, we have investigated a correlation with normalized collisionality $\nu^*$ defined as:
\begin{equation*}
\nu^* = \frac{\nu_{ei}}{\epsilon \times \omega_b}=2.58 \times 10^4 \times \frac{q_{95} \times (R/\textrm{m})^{2.5} \times (n_e/10^{19}\textrm{m}^{-3})^3 \times (V/\textrm{m}^3)^2}{(a/\textrm{m})^{1.5} \times (W_{mhd}/\textrm{J})^2},
\end{equation*}
where $\nu_{ei}$ is the electron ion collision frequency, $\epsilon$ is the aspect ratio, $\omega_b$ is the bounce frequency, R is the major radius and a is the minor radius. $n_e$ is obtained from the core interferometry channel. Fixed values are used for R (1.65m), a (0.5m), V (14$m^3$) and the chord length of the line of sight of the interferometer inside the confined plasma (1m). Figure \ref{fig:sol_col2} shows that for lower values of $\nu_*$ the distribution of $slt_{max}$ is clearly shifted towards higher values.\\ 
\begin{figure}[ht]
  \centering
  \includegraphics[scale=0.5]{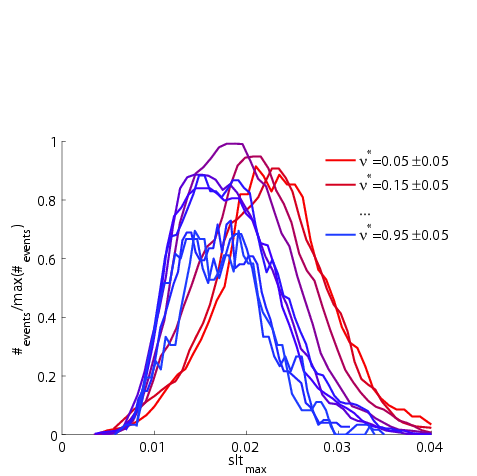}
  \caption{\label{fig:sol_col2} Distribution of \textit{slt} for a variation of intervals of normalized collisionality $\nu^*$: Note that a median filter has been applied on the resulting graphs.}
\end{figure}
The question of whether the observation of SMPs is restricted to a certain ELM type has not been investigated systematically. To take an example: discharge 25764 displays  numerous ELMs with magnetic perturbations of very solitary appearance (see figures \ref{fig:25764_2_elms_sol_comp2} and \ref{fig:25764_mag_mapping3}) with ELM frequencies between 50 and 150Hz. While increasing (decreasing) the input power in this discharge the ELM frequency increases (decreases). In addition, during the investigated phase the discharge is heated with $7.5MW$ by neutral beam injection, which is significantly further above the typical LH-threshold for the parameters of this discharge ($P_{LH}\sim1.6MW$ \cite{RYTER09A}) than type III ELMs are usually observed. Furthermore it is possible to clearly discriminate between ELM event and inter-ELM phase, which is usually not the case for type II ELMs. On the basis of these observations and the criteria given in \cite{ZOHM96A} we infer that the ELMs in discharge 25764 are of type I. 
\subsection{Comparison with magnetic fluctuations evoked by passing mono- or bi-polar current filaments}
\label{subsec:ELM_bi_mono}
The objective of this section is to deduce from the magnetic field perturbation the structure of the current perturbation evoking it. We use a forward modeling approach, comparing measured trajectories of the time derived magnetic field with ones resulting from basic current models. We will first apply this method to the edge snake \cite{SOMMER11A}, where it results in a very clear and simple answer. After that we will return to the ELM, where the situation seams to be less uniform.
\subsubsection{Application to the edge snake (excursion)}
\begin{figure}[ht]
  \centering
  \includegraphics[scale=0.4]{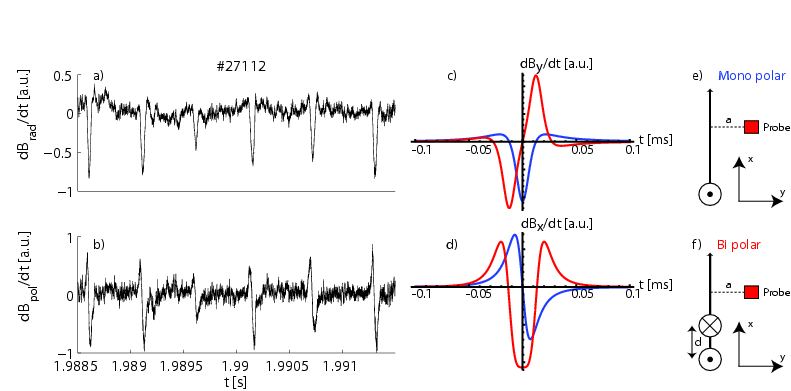}
  \caption{\label{fig:edge_snake_exp_mod} Time derivative of the magnetic field components in the poloidal plane perpendicular (a) and parallel (b) to the flux surface in discharge 27112. Simulated evolution of the time derived magnetic field components on the y-direction (c) and x-direction (d) for the mono-polar (blue) respectively bi-polar (red) configuration. Setup of the mono-polar (e) and bi-polar (f) case. Parameters: Velocity in x-direction 6km/s, minimal distance a=0.075m, separation of poles d=0.15m}
\end{figure}

To measure the magnetic perturbations associated with edge snakes, a triple axis magnetic probe \cite{SCHMID08A} mounted on the mid plane manipulator is employed. The probe measures the time derivative of the magnetic field components in the horizontal and vertical direction within the poloidal plane and perpendicular to these. The time derived magnetic field components in the poloidal plane parallel ($dB_{pol}/dt$) and perpendicular ($dB_{rad}/dt$) to the flux surface have been reconstructed as illustrated in figures \ref{fig:edge_snake_exp_mod}a) and b). \footnote{Identical calibration for the two raw signal components parallel to the poloidal plane has been assumed.} While the large periodic excursions have dominant odd symmetry for $dB_{pol}/dt$, they have a clearly dominant even symmetry for $dB_{rad}/dt$.\\
To infer information about the current perturbation leading to the observed magnetic perturbations the experimental data is compared with results from very basic models. We assume a mono-polar current filament (figure \ref{fig:edge_snake_exp_mod}e) respectively two bi-polar current filaments, which are offset in the x-direction (figure \ref{fig:edge_snake_exp_mod}f). Furthermore, all filaments are parallel to the z-direction and move in the x-direction passing the probe with a finite minimal distance $a$. Figure \ref{fig:edge_snake_exp_mod} shows the qualitative evolution of the time derived magnetic field components in the x-direction (c) and y-direction (d) for the mono-polar (blue) respectively bi-polar (red) configuration.\\
The edge snake is radially located on a rational surface and rotates in the electron diamagnetic drift direction \cite{SOMMER11A}. Thus its main propagation direction is poloidal. Therefore in the comparison between experiment and model the poloidal (radial) direction has to be identified with the x-direction (y-direction). Comparing the symmetries leads to the conclusion that, for the edge snake the bi-polar model is inconsistent with the measurements while the mono-polar model is consistent.
\subsubsection{Application to SMPs}
In comparison to the edge snake the magnetic signature of SMPs has a considerably larger range of possible shapes. Therefore we analyze distributions of some indicators constructed on the basis of the extent ($\dot{B}_{max}, \dot{B}_{min}$) and timing ($t_{max}, t_{min}$) of the maximum and minimum of $\dot{B}_{rad}$ after subtraction of the mean value in an interval of 1ms centered at $t_{I,div,ons}$. The analysis is carried out with data from one magnetic probe (B31-02) (location displayed in figure \ref{fig:mapping_trajectories}). The angle between the measurement direction (perpendicular to probe) and the normal vector on the separatrix in the vicinity of the probe is of the order of a few degrees and is not compensated. The ELM set described above has been confined by further criteria: $slt > 0.03$, $\dot{B}_{max}-\dot{B}_{min} > 3\times std(\dot{B}_{rad})$, $|t_{max}-t_{min}|< 0.1ms$. The first and second criteria are chosen to allow only for very isolated events. The third one ensures that the main peak and dip are close enough to belong to the same passing structure.\\
\begin{figure}[ht]
  \centering
  \includegraphics[scale=0.35]{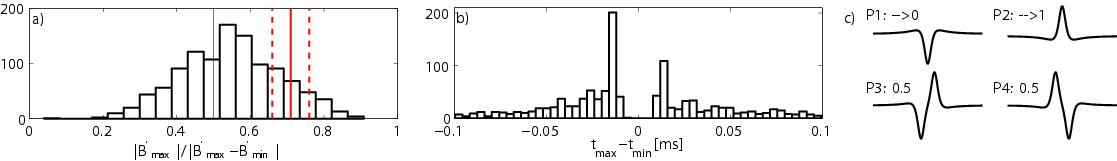}
  \caption{\label{fig:peak_analysis} Histograms of quantities characterizing main peaks and dips of time derived radial magnetic field measured by a magnetic probe at the outer mid plane for selected ELMs (criteria described in the text): a) $|\dot{B}_{max}|/|\dot{B}_{max}-\dot{B}_{min}|$ (red vertical lines indicate mean value and standard deviation of the corresponding quantity for the edge snake displayed in figure \ref{fig:edge_snake_exp_mod}), b) $t_{max}-t_{min}$. b) Prototype waveforms including values for $|\dot{B}_{max}|/|\dot{B}_{max}-\dot{B}_{min}|$: P1, P2: Mono-polar and P3, P4:Bi-polar. }
\end{figure}
We investigate the shape of the perturbations guided by the four prototype waveforms illustrated in figure \ref{fig:peak_analysis}c). To determine if the investigated waveforms have more similarity to an even (P1 and P2) or odd (P3 and P4) symmetry graph we balance $|\dot{B}_{max}|$ and $|\dot{B}_{min}|$. Figure \ref{fig:peak_analysis}a) shows that the histogram of $|\dot{B}_{max}|/|\dot{B}_{max}-\dot{B}_{min}|$ has its maximum close to 0.5 (e.g. $|\dot{B}_{max}|=|\dot{B}_{min}|$) with a slight tendency towards higher values (e.g. $|\dot{B}_{max}| > |\dot{B}_{min}|$). It should be noted that applying the same analysis to unfiltered $\dot{B}_{rad}$ data leads to a qualitatively identical and quantitatively very similar result. The red vertical lines in figure \ref{fig:peak_analysis}a) indicate mean value and standard deviation of the corresponding quantity for the edge snake displayed in figure \ref{fig:edge_snake_exp_mod}. These values are far off 0 and 1, which are the ideal limits for the even symmetry prototypes. However they correspond to $|\dot{B}_{max}|$ being in average higher than $|\dot{B}_{min}|$ by a factor of about 2.3.\\
Thus only a subset of the investigated ELMs at the wings of the distribution shown in figure \ref{fig:peak_analysis}a) can have clear even symmetry. For the rest of ELMs the observed waveforms are incompatible with a mono-polar current filament moving in the poloidal direction. A visual inspection of some of the investigated trajectories confirms this result. Here a significant fraction of trajectories with clear odd symmetry has been observed. However, a number of examples with a short sequence of peaks or dips of similar extent to $\dot{B}_{max}$ respectively $\dot{B}_{min}$ has been observed as well. The latter cases are compatible with a multi-polar (both polarities in alternation) current model.\\
Vianello \cite{VIANELLO11A} has investigated magnetic signals acquired by a triple axis magnetic probe mounted on the mid plane manipulator during ELMs in ASDEX Upgrade. In this work it was inferred that the investigated perturbations are consistent with mono-polar current filaments propagating in the SOL. The fact that this is a different picture to the one we observe for SMPs suggests that two different features have been observed. For instance the ELM filament analyzed most in \cite{VIANELLO11A} occurs about 1ms after the ELM onset, while SMPs appear close ($\pm100\mu s$) to the onset of the pedestal erosion.\\
In figure \ref{fig:peak_analysis}b) a histogram of $t_{max}-t_{min}$ is plotted. Positive (fig. \ref{fig:peak_analysis}c) P3) and negative (fig. \ref{fig:peak_analysis}c) P4) values of this quantity are of the same order of magnitude. The gap at $t_{max}-t_{min}=0$ corresponds to the minimal distance between peaks and dips.\footnote{If the analysis is performed without low pass filtering $\dot{B}_{rad}$ the gap narrows but persists.} The peaks of the distributions are located at $\pm(10 \textrm{ to } 20)\mu s$. Assuming two field aligned bi-polar current filaments at the separatrix rotating with a typical poloidal SMP velocity ($v_{SMP,pol}=10$km/s) these times would correspond to a perpendicular distance between the filaments of 10cm to 20cm. This compares to poloidal filament extensions of 5-8cm observed by Thomson Scattering \cite{KURZAN05A} and visible camera observation \cite{KIRK05A}.
\subsection{Parallel structure and dynamics of SMPs}
\label{subsec:par_struc}
\begin{figure}[ht]
  \centering
  \includegraphics[scale=0.6]{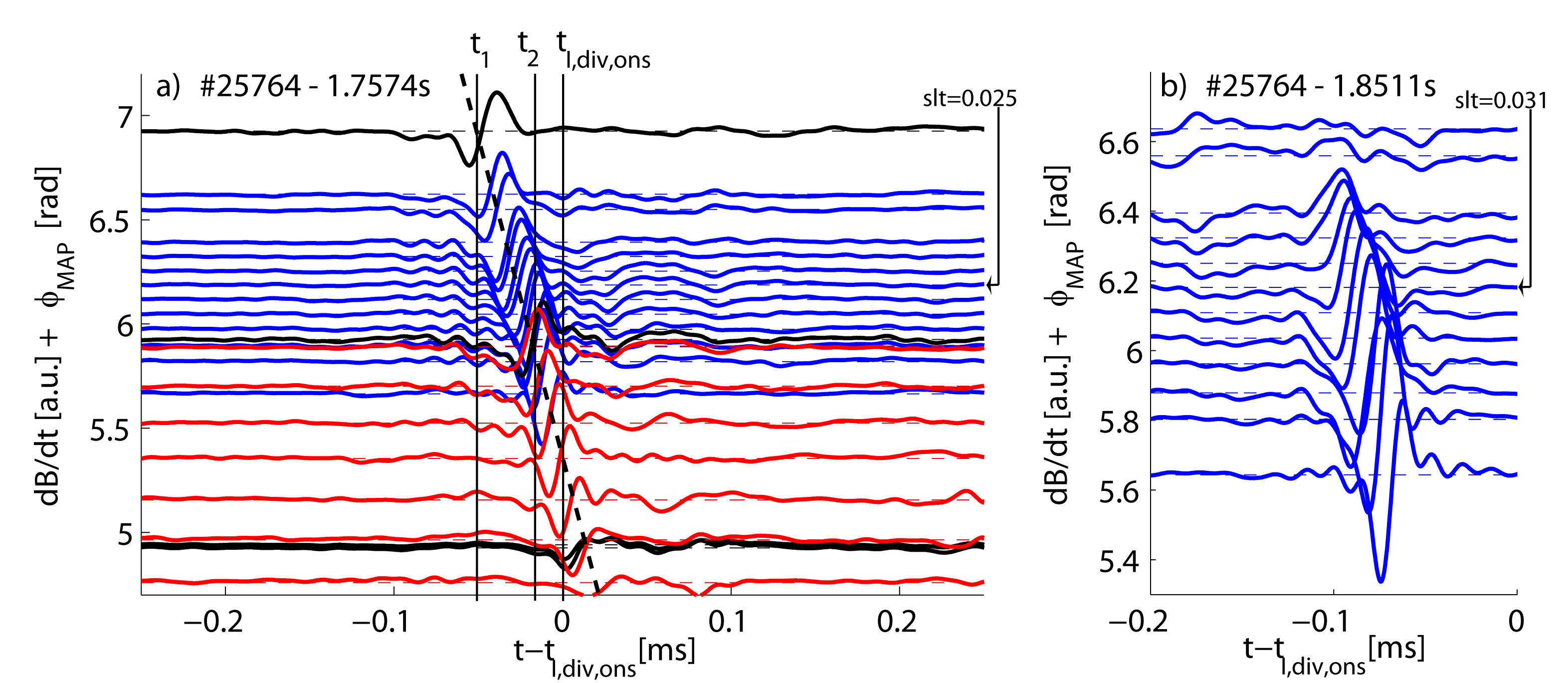}
  \caption{\label{fig:25764_mag_mapping3} Time derived radial magnetic field measured by various pick up coils (color code indicated in figure \ref{fig:mapping_trajectories}) displaced by the toroidal mapping target angle during ELMs in discharge 25764: a) ELM at 1.757s: Dashed lines represent propagation of perturbation (see text). Solid vertical lines mark times $t_1$, $t_2$ and $t_{I,div,ons}$ (see figure \ref{fig:mag_profs}). b) ELM at 1.851s. \textit{slt} values for selected probes are quoted.}
\end{figure}

In order to investigate spatial structure and evolution of SMPs diagnostic mapping as introduced in \ref{subsubsec:diagmap} is applied. Figure \ref{fig:25764_mag_mapping3}a) shows signals from various pick up coils measuring the radial magnetic field fluctuations during an ELM with a very peaked magnetic perturbation in discharge 25764. The signals are displaced by the toroidal mapping target angle, where the mapping procedure has been carried out on the $\rho_{pol}=0.95$ flux surface. Within the traces from the high resolution array (blue) a very solitary magnetic perturbation can be observed. A similar SMP can be observed as well on most channels of the other arrays. On the channels with lower toroidal mapping target angles the peak intensity is reduced. In the space spanned by time and toroidal mapping target angle the points of a certain phase (e.g. zero crossing) of all arrays are located very close to one straight line (black dashed line in figure \ref{fig:25764_mag_mapping3}a). This is consistent with a structure causing the magnetic perturbation, for which the assumption $k_{||}<<k_\perp$ is valid and which is rotating with constant velocity in the toroidal direction.\\
In the example displayed in figure \ref{fig:25764_mag_mapping3}a) the line of constant phase represents a propagation towards lower toroidal mapping target angle, which corresponds to the electron diamagnetic drift direction when mapped in the poloidal direction. In a similar way we have investigated 13 ELMs in this discharge with best traceability of phase. Multiplying the obtained angular velocities of the SMP by a radius of 2m leads to an apparent toroidal velocity\footnote{The measurements described here cannot clarify if the field aligned perturbation is propagating toroidally or poloidally or both. Therefore we define an apparent toroidal SMP velocity $v_{SMP,tor}$ and an apparent perpendicular SMP velocity $v_{SMP,perp}$ as the propagation velocity of the intersection of a field line parallel to the SMP with a line with $\theta=const.$ respectively $parallel\ coordinate=const.$ The conversion between these velocities is carried out on the basis of the approximation that the field line is locally straight: $v_{SMP,perp}/v_{SMP,tor}=-B_\theta/\sqrt{B_\phi^2+B_\theta^2}$} of the SMP of $48\pm11$km/s corresponding to an apparent perpendicular velocity of $10\pm2$km/s (LFS). While all of these SMPs propagate in the electron diamagnetic drift direction, there are isolated cases of SMPs moving in the ion diamagnetic drift direction.\\
The toroidal propagation of the onset location of magnetic activity corresponding to the electron diamagnetic drift direction has been reported earlier for COMPASS-D, JET and AUG \cite{VALOVIC94A,BECOULET03A,BOBKOV04A,NEUHAUSER08A}. In \cite{VALOVIC94A} and \cite{NEUHAUSER08A} single peaks of time integrated components of $dB/dt$ have been followed in this direction.\\ 
As the SMP is observed at positions with different toroidal mapping target angles at different times, it is possible to track the evolution of its shape. Next to perturbations with virtually identical trajectories on all channels two classes of temporal variations are observed. Firstly some SMPs are found to grow or decay. We compare time and value of the peak of $dB_{rad}/dt$ of all probes for the fast growing SMP displayed in figure \ref{fig:25764_mag_mapping3}b). This peak value is growing within about $10\mu s$ by a factor of 4 (growth time $\sim7\mu s$). After this first phase saturation can be observed.\\
Also it is possible that the shape of the time trace is morphing when changing the toroidal mapping target angle $\phi_{map}$ (i.e. moving from probe to probe). In the example displayed in figure \ref{fig:25764_mag_mapping3}b) a transfer from a shape with dominant even symmetry (e.g. a peak only) into a shape with dominant odd symmetry (e.g. a dip followed by a peak) can be seen.
\subsection{Timing of SMP appearance}
\begin{figure}[ht]
  \centering
  \includegraphics[scale=0.4]{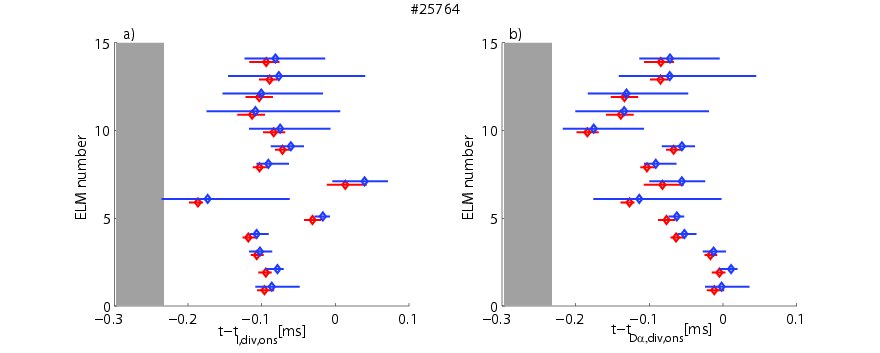}
  \caption{\label{fig:smp_timing} Timing of SMP observation on probes of high resolution poloidal array in discharge 25764 from 1.5s to 2.7s: The bars correspond to $[\mu-\sigma,\mu+\sigma]$, where $\mu$ and $\sigma$ are mean and standard deviation of the distribution of $t_{min}-t_{ref}$ (red) and $t_{max}-t_{ref}$ (blue) for an individual ELM. For $t_{ref}$ a) the onset time of the current in the outer divertor $t_{I,div,ons}$ and b) the onset time of $D_\alpha$ radiation observed in the outer divertor $t_{D\alpha,div,ons}$ has been used. Diamonds indicate median value of the distribution of $t_{min}-t_{ref}$ and $t_{max}-t_{ref}$.}
\end{figure}

A first impression on the timing of SMP observation with magnetic probes relative to the time of the onset of divertor current or $D_\alpha$ radiation in the outer divertor (reference time) can be obtained from figure \ref{fig:26764_ELM_dyn3} c), which shows corresponding time traces for a single ELM. Furthermore for a set of magnetic probes the times $t_{min}(i_{probe})$ and $t_{max}(i_{probe})$ are evaluated, which correspond to the maximum or minimum of the time derived radial magnetic field (figure \ref{fig:smp_timing}). Here 14 ELMs with very clear SMPs in discharge 25764 are
For both choices of reference time most SMPs are observed on the magnetics already up to 0.2ms before $t_{ref}$ and rarely after $t_{ref}$. Thus SMPs appear clearly later than typical coherent ELM precursors (type I and type III) are observed at AUG and JET \cite{KASS98A,PEREZ04B}. To further assess this durations two aspects have to be considered:\\
1) Toroidal rotation: The measurements for both divertor current and divertor $D_\alpha$ radiation are taken at a single toroidal position each. Thus these reference times do not necessarily correspond to the time of the first effect at the divertor. There might be a lag between the detection of SMPs and effects in the outer divertor due to rotation of SMPs to a position, which is connected along field lines to the observed area in the divertor. The apparent toroidal rotation velocity of SMPs obtained in \ref{subsec:par_struc} corresponds to $t_{tor,rot}=0.26ms$ for a full rotation.\\
2) Parallel transport to the divertor: In \cite{Eich09A} for an AUG discharge in upper single null configuration with plasma parameters comparable to discharge 26764 information on typical time scales of parallel transport form the outer mid plane to the outer divertor has been obtained. In this study a good fit of divertor infrared thermography data with results from a free streaming ion model has been obtained. Using this model and the fit results gives for the time between the start of the parallel propagation at the mid plane and the flux at the divertor exceeding $10\%$ to $20\%$  of the peak value $t_{par,transp}=215\mu s$ to $238 \mu s$.\\
In summary peak SMP activity is observed usually later than the start of enhanced parallel transport to the divertor $t_{st,par,transp}=t_{ref}-a \cdot \Delta t_{tor,rot}-\Delta t_{par,transp},\ 0>a>1$ (qualitatively described by the gray shaded region in figure \ref{fig:smp_timing}). The onset of pedestal erosion (figure \ref{fig:26764_ELM_dyn3}) is observed usually less than $100\mu s$ before or after peak SMP activity is observed.
\subsection{Radial location of SMPs}
\label{subsec:rad_loc_smp}
\begin{figure}[ht]
  \centering
  \includegraphics[scale=0.4]{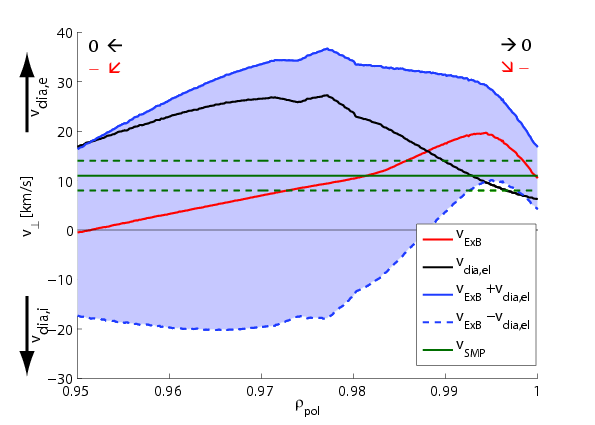}
  \caption{\label{fig:26324_vel_comp} Comparison of perpendicular velocities for 2.5s to 3.5s in discharge 26324: $v_{E \times B}$ (red), $v_{el,dia}$ (black), $\lbrack v_{E \times B}-v_{el,dia};v_{E \times B}+v_{el,dia} \rbrack$ (shaded region) and $v_{SMP,}$ (solid green with interval of one standard deviation in dashed green).}
\end{figure}

Information on the radial location of current perturbations causing SMPs is a key in the assessment of the general role of SMPs in the ELM dynamics. In order to obtain this information we compare the perpendicular rotation of SMPs to the range of perpendicular rotation velocities of a mode (ideal or resistive) expected on the basis of theoretical considerations. In \cite{UZAWA10A} rotation of magnetic islands have been investigated with linear and non-linear calculations. The range of velocities found was well contained within $\lbrack v_{E \times B}-v_{el,dia};v_{E \times B}+v_{el,dia} \rbrack$. To be on the safe side we use the latter interval as estimation of the possible velocities of a mode. Figure \ref{fig:26324_vel_comp} illustrates profiles of $v_{E \times B}$, $v_{el,dia}$, $v_{E \times B}-v_{el,dia}$ and $v_{E \times B}+v_{el,dia}$ for the time interval [2.5s,3.5s] in discharge 26324. The displayed radial range ($0.95\leq\rho_{pol}\leq1.0$) corresponds to the overlap of data availability for the main employed diagnostics (charge exchange recombination spectroscopy and Thomson scattering).\\ 
SMPs can be observed on the magnetic signals during many ELMs in the investigated time interval. However for a subset of them a significant signature from core-MHD-activity is superimposed, which is hampering the analysis. For the 9 purest SMPs positions of constant phase over all signals are traced (compare black dashed line in figure \ref{fig:25764_mag_mapping3}). From the slope of this trajectory in combination with major radius and local field line inclination we obtain an apparent perpendicular rotation velocity of the SMPs $v_{SMP,perp}=11\pm3$km/s (see \ref{subsec:par_struc}). This velocity is shown as well in figure \ref{fig:26324_vel_comp}.\\
For the entire radial range displayed in figure \ref{fig:26324_vel_comp} $v_{SMP,perp}$ is inside $\lbrack v_{E \times B}-v_{el,dia};v_{E \times B}+v_{el,dia} \rbrack$. Due to the profile of the pressure gradient $v_{el,dia}$ will approach zero, when leaving this radial interval into either direction. As well it is known that $v_{E \times B}$ will rapidly turn to negative values outside the illustrated radial range \cite{SCHIRMER06A}. In summary the region, where $v_{SMP,perp}$ is contained within the theoretically predicted interval, extends from the separatrix to the pedestal top or slightly further inside. This finding is in line with results from an earlier detailed comparison of data from ASDEX Upgrade and forward modeling \cite{SCHMID07A,SCHMID08A,NEUHAUSER08A}, which came to the conclusion that the magnetic ELM signature is dominated by structures inside the separatrix.\\
\subsection{Perpendicular structure of SMPs}
\begin{figure}[tb]
  \centering
  \includegraphics[scale=0.4]{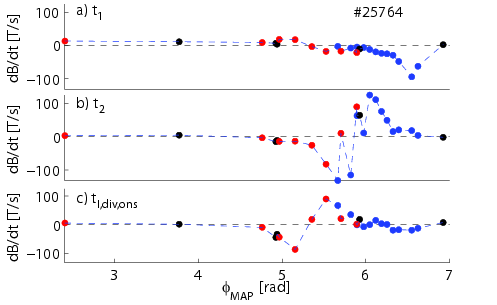}
  \caption{\label{fig:mag_profs} Profiles of time derived radial magnetic field versus toroidal mapping target angle for the three points in time $t_1$, $t_2$ and $t_{I,div,ons}$ marked in figure \ref{fig:25764_mag_mapping3} (color code indicated in figure \ref{fig:mapping_trajectories}).}
\end{figure}

As stated above our observations are consistent with field aligned current structures evoking the SMPs. We now investigate the number $n_{dom}$ of dominant peaks per toroidal turn for the SMP in figure \ref{fig:25764_mag_mapping3}a). The resolution limit in terms of $n_{dom}$ is determined by the angle spanned by the toroidal mapping target angles of the coil positions. Using only sensors in one poloidal plane (e.g. red graphs in fig. \ref{fig:25764_mag_mapping3}a)) enables to resolve $n_{dom}=6$ and higher. Benefiting from the mapping technique allows to use a set of coils with toroidal mapping target angles covering in appropriate density a higher part of the circumference. With this set $n_{dom}=3$ and higher could be resolved.\\
As stated above the black dashed line in figure \ref{fig:25764_mag_mapping3}a) illustrates the trajectory of the central zero crossing of the SMP in the $t-\Phi_{map}$-space. At the time $t_1$ this trajectory passes the upper end of the interval covered by the  $\Phi_{map}$-values. Figure \ref{fig:mag_profs} illustrates profiles of time derived radial magnetic field versus toroidal mapping target angle for the three points in time $t_1$, $t_2$ and $t_{I,div,ons}$ marked in figure \ref{fig:25764_mag_mapping3}. As for the entire displayed time interval at the time $t=t_1$ only one dominant dip respectively peak can be observed in the entire $\Phi_{map}$-interval.\footnote{In figure \ref{fig:25764_mag_mapping3}a) besides this dominant peak a weaker structure can be observed. Comparison with the raw data shows that this structure is due to the applied low pass filter in combination with the strong dominant peak.} However, if there would be 3 or more dominant, equally spaced dips respectively peaks per toroidal turn a further peak should be observed at time $t_1$. Hence $n_{dom}$ is equal to 1 or 2.\\
\begin{figure}[ht]
  \centering
  \includegraphics[scale=0.4]{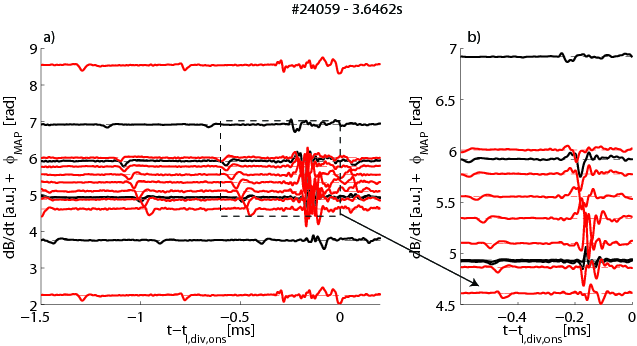}
  \caption{\label{fig:examples_n_discussion} a) Time derived radial magnetic field measured by various pick up coils (color code indicated in figure \ref{fig:mapping_trajectories}) displaced by the toroidal mapping target angle prior to and during an ELM in discharge 24059 at 3.6462s. An identical time trace measured by the filament probe is shown on top and bottom with a relative vertical displacement of $2\pi$. b) Close up of a) with lower signal amplification.}
\end{figure}
More information on the question on the number of dominant SMP peaks per toroidal turn can be obtained studying the relationship of edge snakes and ELMs. Figure \ref{fig:examples_n_discussion} shows an ELM and about 1.5ms of the preceding phase in discharge 24059. An SMP can be observed about 0.2ms before $t_{I,div,ons}$. Prior to this the characteristic dips of an edge snake can be observed. Tracing the dips of the edge snake over several toroidal turns finally leads exactly into the trajectory of the SMP. The velocity of the edge snake perturbation is clearly lower than the one of the SMP during the ELM phase. Similar behavior is observed on a regular basis in the time interval from 3.5s to 3.9s in this discharge. Here 25 cases respectively 1 case are observed of edge snake and subsequent SMP on virtually the same respectively clearly different trajectories.\\
This observation suggests that the edge snake is developing into the structure causing the SMP during the ELM. The edge snake has been reported to have a fundamental toroidal mode number of 1 \cite{SOMMER11A}. As can be seen this is as well the case for the edge snake in figure \ref{fig:examples_n_discussion}. Due to the fast transition into the SMP and the relatively short life time of the SMP it seems natural to assume that the number of dominant peaks per toroidal turn is identical for the edge snake and the SMP during the ELM. This would imply that $n_{dom}$ is equal to one for the SMP during this ELM.\\
In summary there is clear evidence that the number of dominant peaks per toroidal turn for the investigated SMPs is 1 or 2. This result is of special significance when compared to results from linear stability analysis \cite{SNYDER09A}. Here a toroidal mode number $n$ of 1 or 2 is even below mode numbers of the most unstable modes for typical kink/peeling modes ($n \sim 3-6$) and clearly below the ones for typical peeling-ballooning modes ($n \sim 5-20$).\\
\subsection{Temperature perturbation associated with SMPs}
\begin{figure}[ht]
  \centering
  \includegraphics[scale=0.4]{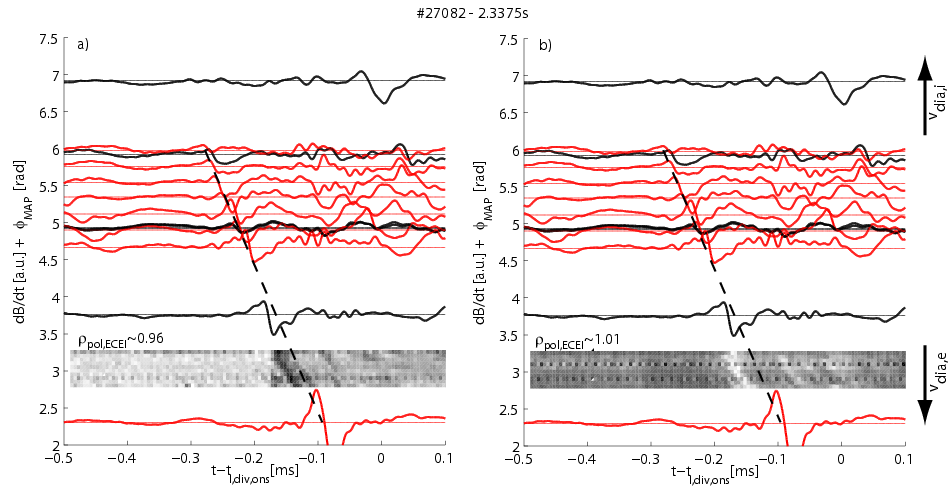}
  \caption{\label{fig:27082_2_3375_mag_ecei_3_5} Correlation of time derived radial magnetic field measured by two arrays (trajectories: color code indicated in figure \ref{fig:mapping_trajectories}) and uncalibrated data from ECEI (images) in discharge 27082: In both plots magnetic data are identical. ECEI data are taken from 11 channels measuring at different vertical but identical radial positions (black to white: cool to hot). These positions correspond roughly to the same poloidal flux: a) $\rho_{pol}\sim0.96$, b)  $\rho_{pol}\sim1.01$. Dashed lines represent propagation of perturbation (see text).}
\end{figure}

Electron Cyclotron Emission Spectroscopy (ECE) is measuring the intensities of millimeter waves at defined frequencies. For optically thick plasmas these intensities can be assumed to be proportional to the electron temperature. This condition is usually fulfilled at locations inside the steep gradient region. In \cite{BOOM11A} a profile of the optical thickness $\tau$ has been calculated for an ASDEX Upgrade H-mode plasma on the basis of the profiles of electron density and temperature. The transition ($\tau = 3$) from optical thick to thin is well coinciding with the location of the separatrix. For optically thin conditions solving the radiation transport equation would be necessary in order to infer reliable values of the electron temperature.\\
ASDEX Upgrade is now equipped with an ECEI diagnostics \cite{BOOM11A}, which is measuring radiation temperature in X2-mode on a rectangular grid of positions located in one poloidal plane (27082: R=2.07m to 2.17m, Z=-0.10m to 0.25m). Figure \ref{fig:27082_2_3375_mag_ecei_3_5} correlates the time derived radial magnetic field and uncalibrated data from ECEI. The magnetic traces are identical in both subplots. Due to the application of diagnostic mapping it is possible to compare the propagation of the SMP with the evolution of the temporal-spatial evolution of the radiation temperature pattern. In both subplots a dashed line marks the trajectory of the central zero-crossing of an SMP in the $t-\Phi_{map}$-space. Parallel and near to this line in figure \ref{fig:27082_2_3375_mag_ecei_3_5}a), which is showing ECEI data from inside the separatrix, a track of decreased radiation temperature can be observed. Assuming optical thickness this could be interpreted as a temporary electron temperature reduction propagating in $\theta-\Phi$-space on the same field line as the SMP. However we can not exclude that the reduced radiation temperature is caused by a local density enhancement in excess of the critical density for X2-cutoff located radially outside the probed position.\\
Figure \ref{fig:27082_2_3375_mag_ecei_3_5}b) shows ECEI data from channels with measurement positions just outside or at the separatrix for the same ELM. Here a significant temporary radiation temperature increase parallel and close to the dashed line is observed. We recall that the plasma is of marginal optical depth in this region. Thus it is not clear, if the observed perturbation of radiation temperature is associated with an equivalent perturbation of electron temperature or if it is caused by a nonlocal effect. The described behavior at and inside the separatrix can not be observed for each ELM, but rather occasionally. In many cases the phenomenology only inside or only outside the separatrix can be seen.\\
In summary in the presented example a decrease of the electron temperature slightly inside the separatrix in combination with an increase of radiation temperature slightly outside or at the separatrix correlated with the SMP has been observed. Thus there might be a flattening of the electron temperature profile correlated to the SMP. In principle this would be compatible with an island structure of the SMP.\footnote{Note that the observed SMP onset dynamics (see \ref{subsec:par_struc}) does not support such a hypothesis!}\\
In \cite{BOOM11A} a coherent mode with around 18 peaks per toroidal turn appearing a few hundreds of $\mu s$ before $t_{I,div,ons}$ has been reported. It is important to note that for the ELM shown in figure \ref{fig:27082_2_3375_mag_ecei_3_5} such a mode has not been observed.
\section{Correlation of SMPs to ELM filaments}
\label{sec:gpi}
\begin{figure}[ht]
  \centering
  \includegraphics[scale=0.5]{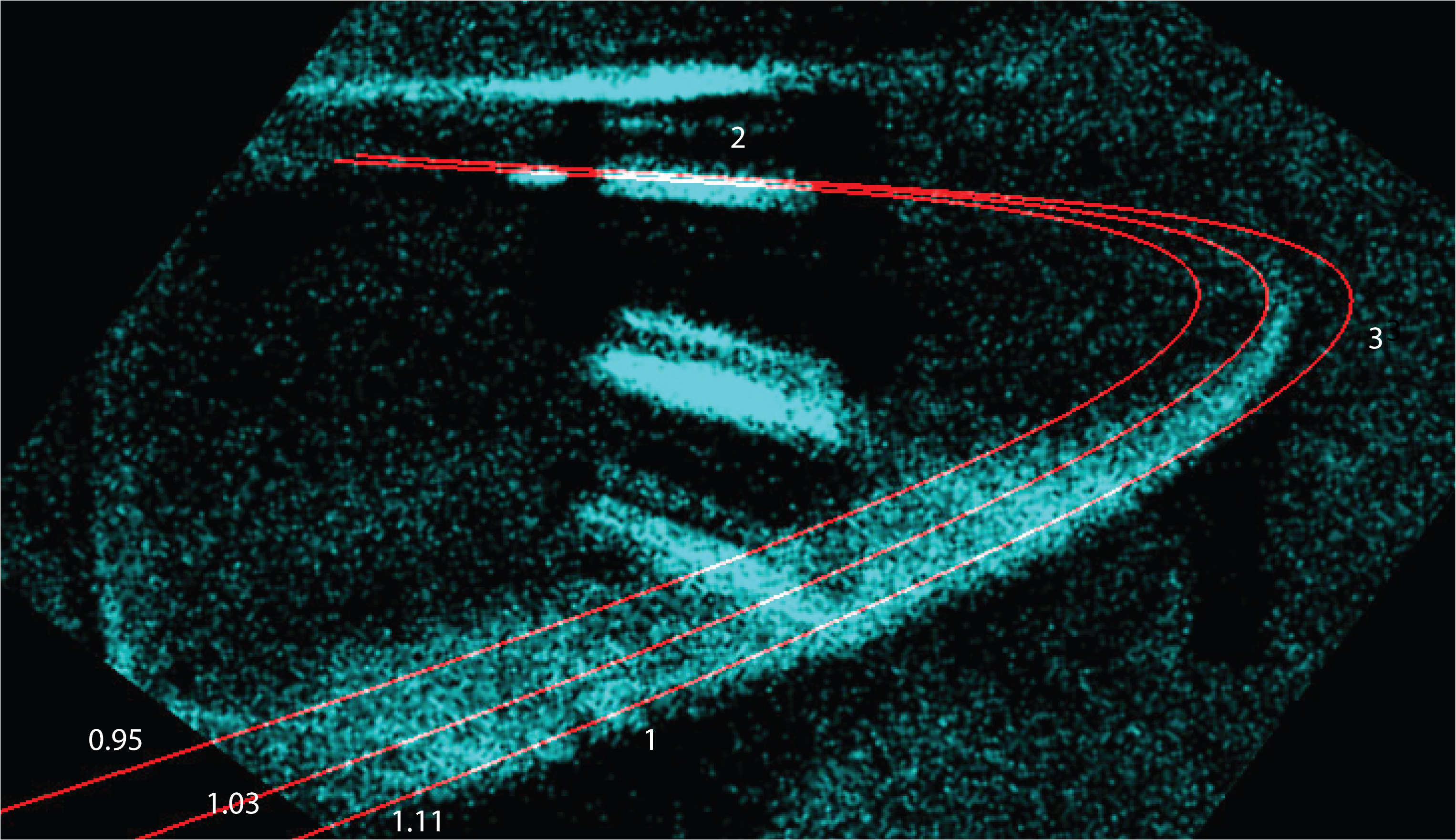}
  \caption{\label{fig:long_fils} ELM radiation filament in discharge 26299 observed by fast framing camera. Field lines on flux surfaces with $\rho_{pol}$ = 0.95, 1.03 and 1.11 are displayed in red.}
\end{figure}

\begin{figure}[ht]
  \centering
  \includegraphics[scale=0.4]{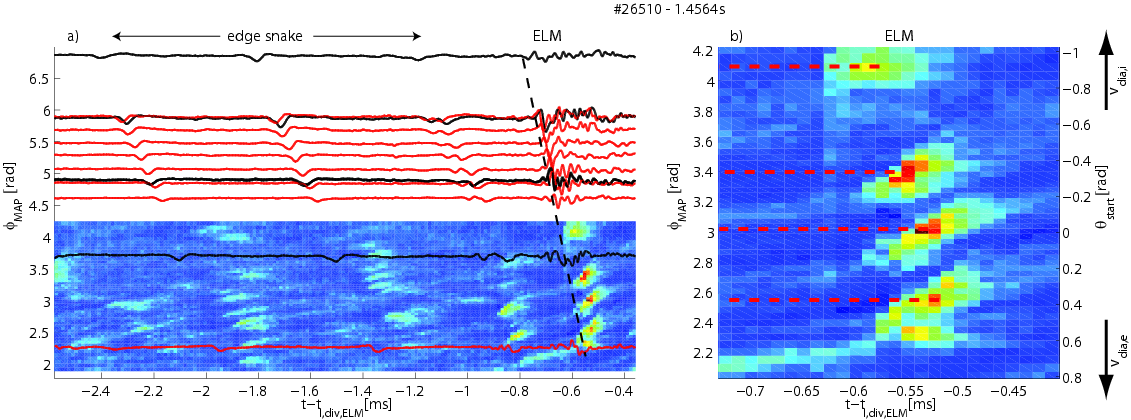}
  \caption{\label{fig:26510_1_4564_mag_fcam_both2} Evolution of filaments at 1.456s in discharge 26510: a) Correlation of radial magnetic field (trajectories: color code indicated in figure \ref{fig:mapping_trajectories}) and data from fast framing camera processed as described in the text (image) in $t-\Phi_{map}$-space. Dashed line represents propagation of an SMP. b) Close up on processed data from the fast framing camera. Right axis provides the values of $\theta_{start}$. Dashed lines indicate average mapping target angles at the time of the highest intensity.}
\end{figure}

It is of high interest to study the correlation of SMPs and signatures of the non-linear evolution of ELMs. Filamentary structures in the SOL are a feature of the fully developed non-linear phase of ELMs \cite{FUNDAMENSKI07B}. They can be observed with fast framing cameras \cite{KIRK04B} especially well if appropriate gas puff is applied.\\
In the discharges presented here a camera setting integrating 256 x 256 pixels for  $12\mu s$ with a frame rate of 79kframes/s has been used.\footnote{This time resolution constitutes a compromise between collecting sufficient signal and capturing relevant time scales.} In order to visualize radiation patterns associated with ELMs a background subtraction technique is applied in order to remove radiation components varying on slower time scales. From each frame the average of the set of frames starting 10 frames earlier and finishing 10 frames later has been subtracted. \\
ELM filaments are clearly less intensely observed, if the Deuterium fueling is transferred from the main chamber port the camera is viewing at to a port half a toroidal turn apart or to the divertor. If the gas puff is applied from the port in the camera view and a filter allowing for virtually no transmission of $D_\alpha$- or $D_\beta$-light is used, no ELM filaments have been observed. This suggests that radiation from the ELM filaments is Deuterium radiation.\\
We use two tangential views from the same port viewing horizontally in different directions combined with two special gas puff recipes. The first view-gas-setting is leading to the appearance of radiation ELM filaments virtually in the entire view. Figure \ref{fig:long_fils} shows a radiation filament with large parallel extension. Field lines with $\rho_{pol} =$ 0.95, 1.03 and 1.11 are shown as well. The filament is field aligned. It passes regions of high neutral density due to a gas puff in the foreground (1) and background (2), where a number of short radiation stripes can be seen above each other. In this view-gas-setting it is possible to trace the filament as well to the location, where they have the maximum curvature in the projection (3). Comparison with superimposed field lines shows that in this region the filament is clearly localised radially outside the last closed flux surface between $\rho_{pol} =$ 1.03 and 1.11. In the foreground (1) the radiation filaments broadens. It can not be decided if here the origin of the radiation is extending further in the poloidal or radial direction or both. If it is extending radially inwards the separatrix, this could be due to neutrals penetrating radially further inside due to the higher neutral pressure in the vicinity of the gas injection port. This would be also in line with the observation, that filaments are sometimes visible in location (2) one or two frames before they appear in other areas (e.g. (3)).\\
Applying the second view-gas-setting the area, where filaments are observed, corresponds well with the boundaries of the port of about 0.5m width, 1.0m height and 1.0m depth, which the gas reaches before entering the main chamber. Therefore we assume that in the volume in front of this port in a fixed radial position the neutral density is roughly homogeneous. Due to the reduced length of the filaments compared to the other setting it is not possible to discriminate by comparison with field lines if the radial position of the filaments is inside or outside the separatrix. However it seams natural to assume that as well for this view-gas-setting a part of the ELM filament radiation originates from the SOL.\\
In discharge 26510 at 1.456s ELM filaments can be clearly observed in the area of the port, through which the gas puff is carried out. The maximum relative radiation enhancement due to the filaments is $35\%$. We assume typical SOL values for the equilibrium electron density and temperature ($10^{19}m^{-3}, 10eV$) in combination with $D_\alpha$ emission and spatially homogeneous neutral density in the area of the port. Under these conditions the observed increase in radiation could either be evoked by an increase in temperature of about 7eV at constant electron density, a density increase of about $4\times10^{18}m^{-3}$ or a combination of density and temperature increase. The latter is probably the case, as simultaneous increase of SOL electron density and temperature have been observed during ELMs before \cite{MUELLER10A}.\\
In order to capture the spatial-temporal evolution of the field aligned radiation structures a special data processing approach has been developed. A 3D viewer \cite{LUNT10A} has been used to handle the transformation between real space and camera chip. In a first step a chain of analysis points with constant values for $\rho_{pol}$ and $\Phi$ and an equidistant set of poloidal angles $\theta_{start}$ is defined. For each of the analysis points a field aligned chain of points with constant toroidal distances is also localised. Using the projection module all points are projected in the coordinate system of the camera chip. For each poloidal starting position $\theta_{start}$ of the analysis chain and each time the average radiation value over the field aligned chain is calculated. As well for each point on the analysis chain the field line is traced to $\theta=0^\circ$ as described in \ref{subsubsec:diagmap} to obtain the toroidal mapping target angles.\\
Figure \ref{fig:26510_1_4564_mag_fcam_both2}b) shows the obtained representation of the spatial-temporal evolution of the filaments during the ELM quoted above. Four propagating filaments can be identified. Detailed analysis shows that the growth time of the peaks is clearly lower than their decay times. The radiation peaks propagate towards lower $\theta_{start}$ (e.g. downwards on the magnetic low field side). Regarding the magnetic configuration of this discharge (standard ASDEX Upgrade configuration) the filaments are moving in the ion diamagnetic drift direction. This propagation direction of ELM filaments is a general feature which is consistently observed at ASDEX Upgrade for ELMs with low to medium frequencies.\\
We have identified the main trajectories for the three filaments observed in the area $-0.6 < \theta_{start} < 0.8$. The corresponding apparent poloidal angular velocity is in average 2.9krad/s. Multiplying these values by the minor radius gives an apparent poloidal rotation velocity of 1.4km/s \footnote{The apparent perpendicular rotation velocity differ from this value by up to $3\%$.}.\\
Regularly sets of filaments appear within a short time interval as shown in figure \ref{fig:26510_1_4564_mag_fcam_both2}b). Here the average mapping target angles at the time of the highest intensity for the four most intense structures are highlighted by dashed lines. The distance between the two uppermost filaments is nearly twice as high as for other neighbors. If the filament structure is transferred from a pedestal mode structure, one could speculate in this case that one filament between the two top ones did not accelerate radially. Extrapolating average distance in $\Phi_{map}$ for the lower three filaments leads to 15 radiation peaks on the outer mid plane per full toroidal turn.\\
Due to the parametrization with the toroidal mapping target angle the evolution of radial magnetic field perturbations can be compared to that of the filaments as shown in figure \ref{fig:26510_1_4564_mag_fcam_both2}a)\footnote{The evolution of $|I_{pol,sol,out}|$ for this ELM consists in two rise phases separated by a short plateau phase. The automatic recognition of $t_{I,div,ons}$ for this ELM has yielded the onset in the second rise phase. The onset in the first rise phase is about 0.55ms earlier.}. The radiation peaks are clustered in time in several sets. On the magnetic signals an edge snake can be traced from the start of the displayed time window propagating about 3 turns in the toroidal direction. At about -0.6ms as well this edge snake is leading into an SMP, which is propagating with clearly higher velocity. The propagation direction of the edge snake and the SMP is the electron diamagnetic drift direction, thus contra the propagation direction of the radiation peaks.\\
The timing of the onset of the radiation peaks is clearly linked to the trajectory of edge snake and SMP. For the SMP phase in figure \ref{fig:26510_1_4564_mag_fcam_both2}b) a tendency of filaments at lower $\Phi_{map}$ setting on later can be clearly observed. In figure \ref{fig:26510_1_4564_mag_fcam_both2}a) a dashed line following the onset of the SMP has been added. Except for the bottom filament the onset time of the radiation peak is well described by the passing time of this trajectory at a given location. We have found numerous ELMs with sets of filaments setting on one after the other when or short after the same SMP passes their location. During the edge snake phase the radiation features have lower distance in $\Phi_{map}$ and lower velocities than during the ELM phase. There is a lag of the order of 0.1ms between passing of the magnetic perturbation associated with the edge snake and first observation of enhanced radiation in this position.\\
In a discharge pair (26703, 26704) in upper single null configuration with an inversion of the toroidal magnetic field direction SMPs are still propagating in the electron diamagnetic drift direction and radiating filaments are still moving in the ion diamagnetic drift direction as in discharges with a magnetic field direction, which is standard at ASDEX Upgrade. In both discharges a number of examples are observed where a link between passing of a certain location by the SMP and onset of a filament as described above can be identified. 
\section{Summary and discussion}
\label{sec:sum_dis}
In this work a number of aspects of ELM dynamics have been studied on the basis of experimental data. Magnetic perturbations have served as a starting point. In particular \textit{solitary magnetic perturbations} have been identified to occur routinely close ($\pm100\mu s$) to the onset of erosion of the electron temperature and density pedestal. The intensity of the peaking of the time derived radial magnetic field has been described by a new quantity named  \textit{solitariness}. In view of the relatively low values of normalized collisionality $\nu^*$ expected for large scale devices such as ITER \cite{LOARTE03B} compared to ASDEX Upgrade it is of high relevance that the solitariness distribution is pushed towards higher values, when reducing $\nu^*$.\\
Further analysis of these perturbations leads to the insight that - in contrast to the magnetic perturbation associated with the edge snake - SMPs can not be generally evoked by propagating mono-polar current filaments. Rather than this in many cases a rotating structure consisting of two or more current filaments of alternating direction can explain the observed patterns of time derived radial magnetic field. Using mapping techniques developed in the context of this work the SMPs and the current perturbations evoking them have been identified as field aligned. For examples of SMPs with high solitariness it has been shown that the number of dominant peaks per toroidal turn must be lower than 3. Comparison with results from linear stability calculations \cite{SNYDER09A} suggests that the SMP is not a signature of a typical linear peeling-ballooning mode but the signature of a non-linear process.\\
SMPs are commonly rotating in the electron diamagnetic drift direction with typical apparent perpendicular velocities of 10km/s. Comparison of these velocities with the range of expected phase velocities for a mode yields that the current perturbations associated with SMPs are located roughly between separatrix and pedestal top. As well it is observed that SMPs are associated with a temperature perturbation, which is located in the same radial region and which is propagating with the same velocity as the SMP in its close vicinity. In the pedestal region this perturbation is usually a decrease in temperature.\\
SMPs are observed to grow on a time scale of $10 \mu s$. This growth phase is followed by a saturation, which is another indication for a non-linear nature of SMPs. In comparison the theory of explosive ideal MHD-instabilities \cite{WILSON04A} describes a finger-shaped perturbation, which is broadening in the radial direction narrowing in the other perpendicular direction with time. The growth of SMPs is assumed to be as well associated with a local, radially extending perturbation of poloidal flux. Thus the theory presented in \cite{WILSON04A} is compatible with our experimental findings so far, but further analysis is needed to carry out this comparison in detail.\\
SMPs are considered to be features of the non-linear phase of an ELM. Radiation filaments are another signature of ELMs, which we have observed by gas puff imaging. There is an indication that the observed filaments are at least partly located in the SOL. As this requires the filament to have expanded significantly into the SOL, we assume this to happen in the fully developed non-linear phase. Using again the mapping techniques allows to project data from magnetics and fast framing camera into one common frame. Thus it provides the possibility to correlate two signatures of the non-linear ELM phase, one originating at or inside the separatrix, the other originating at least partly in the SOL.\\
Looking at this correlated data the first observation is that in the lab frame SMPs and radiation filaments are propagating in opposite perpendicular directions namely the electron and the ion diamagnetic drift direction. This clearly shows that SMPs and filaments are not the footprint of the same propagating structure. Further this  can be understood regarding radial location and type of the perturbations. The SMP, which is originating from the separatrix or inside, rotates with the local ExB-velocity (in the electron diamagnetic drift direction) plus a certain phase velocity (see section \ref{subsec:rad_loc_smp}). In contrast the radiating filaments can be regarded to rotate with the local ExB-velocity in the SOL, which is directed in the ion diamagnetic drift direction.\\
It is even more remarkable that the onset time of the radiation filaments is often well described by the time when an SMP passes the respective location. One SMP usually passes a number of locations, where filaments are observed to start. This suggests that the structures evoking SMPs alone or in combination with another deviation from the equilibrium provide conditions, under which radiation filaments can propagate radially. Results guiding in a similar direction are reported in \cite{NEUHAUSER08A}, where data measured by magnetic probes and Langmuir probes from ASDEX Upgrade are analyzed. Here an example of an ELM is presented (figure 6), for which magnetic perturbations pass a combined Langmuir magnetics probe mostly at times, at which peaks on the ion saturation current are registered.\\
We regard SMPs as signature of an important aspect in the non-linear ELM evolution. We have collected indication that they have characteristics comparable to a trigger especially for the radial propagation of filaments. This might make them to highly interesting objects in view of the development of ELM control or mitigation. Further experimental investigations and detailed comparison to results from non-linear simulations of ASDEX Upgrade ELMs with JOREK \cite{HUYSMANS07A,HOELZL11A,HOELZL12A} are planned. Eventually the findings related to SMPs …may contribute to a future quantitative understanding of the non-linear ELM evolution.

\section{References}

\bibliographystyle{ppcf_bib}
\bibliography{rrwbib}

\end{document}